\documentclass[fleqn,usenatbib]{mnras}

\usepackage{newtxtext,newtxmath}
\usepackage[T1]{fontenc}
\usepackage{xcolor}
\usepackage[normalem]{ulem}

\usepackage{orcidlink}
\usepackage[normalem]{ulem}

\DeclareRobustCommand{\VAN}[3]{#2}
\let\VANthebibliography\thebibliography
\def\thebibliography{\DeclareRobustCommand{\VAN}[3]{##3}\VANthebibliography}

\usepackage{graphicx}	%
\usepackage{float}
\usepackage{physics}
\usepackage{amsmath}	

\title[GW-Cosmology]{Gravitational Wave Informed Inference of 21-cm Global Signal Parameters}

\author[A. Tiwari et al.]{Avinash Tiwari \orcidlink{0000-0001-7197-8899},$^{1}$\thanks{E-mail: avinash.tiwari@iucaa.in}  
Sajad A. Bhat \orcidlink{0000-0002-6783-1840},$^{1}$
Tirthankar Roy Choudhury \orcidlink{0000-0001-7462-8587},$^{2}$
\newauthor
Susmita Adhikari \orcidlink{0000-0002-0298-4432},$^{3}$
Mukesh Kumar Singh \orcidlink{0000-0001-8081-4888},$^{4}$ and
Shasvath J. Kapadia \orcidlink{0000-0001-5318-1253}$^{1}$
\\
$^{1}$Inter University Centre for Astronomy and Astrophysics, Post Bag 4, Ganeshkhind, Pune - 411007, India\\
$^{2}$National Centre for Radio Astrophysics, Ganeshkhind - Tata Institute of Fundamental Research, Pune - 411007, India\\
$^{3}$Department of Physics, Indian Institute of Science Education and Research (IISER) Pune, Dr. Homi Bhabha Road, Pune 411008, India\\
$^{4}$Gravity Exploration Institute, School of Physics and Astronomy, Cardiff University, Cardiff CF24 3AA, United Kingdom}

\pubyear{2025}

\begin{document}
\label{firstpage}
\pagerange{\pageref{firstpage}--\pageref{lastpage}}
\maketitle

\newcommand{\SB}[1]{{\color{red}\textit{Sajad: #1}}}
\newcommand{\AT}[1]{{\color{orange}\textit{Avinash: #1}}}
\newcommand{\mks}[1]{{\color{brown}\textit{Mukesh: #1}}}

\begin{abstract}
Understanding how and when the first stars and galaxies formed remains one of the central challenges in modern cosmology. These structures emerged during the transition from the Dark Ages to the Cosmic Dawn, a period that remains observationally unconstrained despite strong theoretical progress. During this epoch, neutral hydrogen absorbed a fraction of cosmic microwave background photons through its 21-cm hyperfine transition, producing a 21-cm absorption signal whose evolution encodes the early Universe’s thermal and ionization history. However, extracting the underlying astrophysical parameters from this signal is limited by severe parameter degeneracies, which cannot be resolved without independent observational probes. The next-generation gravitational wave (GW) detectors, such as Cosmic Explorer (CE), will observe binary black hole (BBH) mergers up to very large redshifts and hence will detect a fraction of them formed within the redshift range $\sim 13-25$. The merger rate of these BBHs will depend on the star formation rate density (SFRD) at these redshifts, together with the BBH formation efficiency and a time delay distribution. Therefore, the merger rate of these BBHs can work as a tracer of the SFRD in the redshift range $\sim 13-25$. In this {\it Letter}, we establish a novel multi-messenger framework and present a proof-of-principle concept of how the observations of BBH mergers form next-generation GW detectors can improve the inference of parameters generating the 21-cm cosmic hydrogen signal, and help break degeneracies between them.
\end{abstract}




\section{Introduction}\label{sec: intro}

The emergence of the first luminous sources -- metal-free Population III stars, galaxies, and quasars -- marked the end of the cosmic Dark Ages and initiated the Cosmic Dawn and the Epoch of Reionization \citep[for reviews, see, e.g.,][]{2001PhR...349..125B,2022GReGr..54..102C}. Despite significant theoretical progress, this major phase transition in the history of the Universe remains one of the least constrained epochs in cosmology \citep{2012RPPh...75h6901P}. While ground- and space-based near- and mid-infrared telescopes, such as the recently launched James Webb Space Telescope (JWST), aim to resolve bright individual sources at high redshifts \citep{2025NatAs...9.1134A}, probes of neutral hydrogen in the diffuse intergalactic medium (IGM) offer a complementary and unique window into the ionization and thermal histories of the Universe, as well as the cumulative radiation fields produced by the first stars \citep{1997ApJ...475..429M,FurlanettoOhBriggs2006}.

Of particular interest is the global 21-cm signal, which traces the sky-averaged evolution of the IGM. The observable quantity is the differential brightness temperature, $\delta T_b$, defined as the contrast between the spin temperature of neutral hydrogen ($T_S$) and the Cosmic Microwave Background (CMB) temperature ($T_\gamma$). The evolution of this signal is governed by the interplay between radiative and collisional processes. During the Cosmic Dawn ($z \sim 15-30$), Lyman-$\alpha$ (Ly$\alpha$) photons from the first stars couple $T_S$ to the adiabatically cooled kinetic temperature of the gas ($T_K$) via the Wouthuysen–Field effect \citep{1952AJ.....57R..31W,1958PIRE...46..240F}, producing a pronounced absorption feature against the CMB. Subsequently, X-rays from early stellar remnants heat the IGM; once $T_K$ exceeds $T_\gamma$, the signal transitions from absorption to emission. As a result, the global 21-cm signal encodes rich astrophysical information and depends sensitively on: (i) the star formation rate density (SFRD), $\Psi(z)$; (ii) the Ly$\alpha$ coupling efficiency, $f_\alpha$, determined by the stellar ultraviolet spectra; and (iii) the X-ray heating efficiency, $f_{Xh}$. Consequently, it is the prime target for precision radiometer experiments, such as the Experiment to Detect the Global EoR Signature (EDGES) \citep{2008ApJ...676....1B, Bowman2018}, the Shaped Antenna measurement of the background RAdio Spectrum (SARAS) \citep{ssingh2022}, the Radio Experiment for the Analysis of Cosmic Hydrogen (REACH) \citep{deLeraAcedo2022}, and the Large-aperture Experiment to Detect the Dark Ages (LEDA) \citep{Price2018}. Detecting this signal, however, is extremely challenging due to Galactic foregrounds that are orders of magnitude brighter than the cosmological signal and due to instrumental systematics, leading to current tensions between reported detections \citep[e.g., EDGES,][]{Bowman2018} and null results \citep[e.g., SARAS~3,][]{ssingh2022}.

Although observations of the 21-cm global signal will significantly advance our understanding of the Cosmic Dawn, the signal suffers from strong parameter degeneracies among $f_\alpha$, $f_{Xh}$, and the SFRD. These degeneracies cannot be broken without independent constraints on at least one of these quantities. In the absence of such constraints, we lose our ability to 
\begin{enumerate}
    \item reconstruct a timeline of first-light events, i.e., the timing and duration of the cosmic dawn, because the redshifts of when the Ly$\alpha$ coupling occurred and when the IGM was heated above the CMB become uncertain;
    \item characterize the nature of first luminous sources because $f_{\alpha}$ depends on the UV luminosity and spectra of early stars, and $f_{Xh}$ depends on the population of sources emitting X-ray photons;
    \item reconstruct the thermal history of the neutral IGM because the balance between X-ray heating and adiabatic cooling determines whether the 21-cm signal will be seen in absorption or emission;
    \item determine when reionization started, for example, delayed heating can mimic a late start of reionization, and vice versa.
\end{enumerate}
Precise constraints on $f_\alpha$, $f_{Xh}$, the epoch of X-ray heating, and the SFRD are therefore critical. This motivates the exploration of complementary probes of the high-redshift Universe capable of breaking these degeneracies. One promising approach is to combine galaxy observations from JWST with measurements of the 21-cm global signal to jointly constrain the underlying physical processes \citep{2022MNRAS.515.2901M,2025MNRAS.542.2292D}, although detecting a sufficiently large population of galaxies at such high redshifts remains challenging.

Another, relatively less explored, complementary probe of the intermediate stages of cosmic evolution is provided by gravitational waves (GWs), particularly through next-generation ground-based detectors such as the Cosmic Explorer \citep[CE, ][]{Reitze:2019iox} and the Einstein Telescope \citep[ET, ][]{Hild:2010id}. These facilities, will be sensitive to stellar mass binary black hole (BBH) mergers happening at very large redshifts ~\citep[$z \sim 20-100$, ][]{Hall:2022dik}, depending on their total mass, and hence will observe a fraction of BBHs formed from the stars born between redshifts $\sim 13-25$. Therefore, the merger rate of these BBHs will depend on the SFRD at these redshifts together with the BBH formation efficiency $\eta (z)$ and a time delay distribution $p(t_d)$, i.e., the distribution of times elapsed between formation and merger of BBHs. Applying {\it hierarchical inference}~\citep{Loredo:2004nn, Talbot:2017yur, Fishbach:2018edt, Thrane:2018qnx, Vitale:2018yhm} on a subpopulation of BBHs observed by CE, we can, in principle, constrain the SFRD, $\eta(z)$, and $p(t_d)$ and then use this SFRD as a prior while inferring the parameters of the 21-cm global signal.

In this {\it Letter}, we present a proof-of-principle demonstration of this method and show that it drastically improves the inference of the 21-cm global signal parameters.

\section{Motivation and Method}\label{sec: method}

\subsection{Modelling the 21-cm global signal}

The 21-cm signal is observed as a contrast against the CMB. The observable quantity is the globally-averaged differential brightness temperature $\delta T_b$, which at a redshift $z$ is given by \citep{FurlanettoOhBriggs2006}:
\begin{equation}
    \delta T_b (\nu) \approx \frac{T_S(z) - T_\gamma(z)}{1 + z} \left(1 - \exp \left[-\frac{0.0092 \, (1 + z)^{3/2}}{T_S(z)}\right] \right),
    \label{eq:delta_Tb}
\end{equation}
where $T_\gamma(z)$ and $T_S(z)$ are the CMB temperature and the spin temperature of neutral hydrogen, respectively. We have assumed the IGM 
to be fully neutral, as our focus in this work is on the Cosmic Dawn epoch (redshifts $z \sim 13-25$), prior to the bulk of reionization.

The evolution of the signal is primarily governed by the spin temperature $T_S$, which determines the relative population of the hyperfine energy levels. $T_S$ represents the equilibrium temperature established by the competition between scattering of CMB photons and the scattering of Ly$\alpha$ photons (the Wouthuysen-Field effect). This equilibrium is described by:
\begin{equation}
    T_S^{-1} = \frac{T_\gamma^{-1} + x_\alpha T_K^{-1}}{1 + x_\alpha},
    \label{eq:spin_temp}
\end{equation}
where $T_K$ is the gas kinetic temperature, and $x_\alpha$ quantifies the efficiency of Ly$\alpha$ scattering. In the above, we have ignored collisions as the number densities of free electrons and protons become negligible due to the expansion of the Universe. We have also assumed that the colour temperature of the Ly$\alpha$ radiation field is equal to the gas temperature.

We compute the thermal history of the IGM using the standard cosmological evolution equation for the gas kinetic temperature $T_K$. Including the effects of adiabatic expansion and X-ray heating, the evolution is given by:
\begin{equation}
    \frac{dT_K}{dz} = \frac{2 T_K}{1+z} - \frac{2}{3 k_B n_H H(z) (1+z)} \epsilon_X(z),
    \label{eq:TK_evolution}
\end{equation}
where $H(z)$ is the Hubble parameter, $n_H$ is the hydrogen number density, and $\epsilon_X$ is the X-ray heating rate density. The first term on the right-hand side represents adiabatic cooling due to cosmic expansion, while the second term represents heating by X-ray sources. We model the X-ray heating rate density as being proportional to the SFRD $\Psi(z)$ \citep{ChatterjeeDayalChoudhuryHutter2019,ChatterjeeDayalChoudhurySchneider2020,ChatterjeeChoudhury2023}:
\begin{equation}
\frac{\epsilon_X(z)}{\mathrm{erg}~ \mathrm{s}^{-1} \mathrm{Mpc}^{-3}} = 3.4 \times 10^{40} \, f_{Xh} \frac{\Psi(z)}{M_\odot \mathrm{Mpc}^{-3} \mathrm{yr}^{-1}},
\end{equation}
where $f_{Xh}$ is the heating efficiency parameter. This parameter normalizes our model to account for uncertainties in the X-ray properties of high-redshift galaxies compared to local observations. Physically, it combines the X-ray production efficiency per unit SFR and the fraction of X-ray energy effectively deposited as heat in the IGM.

The Ly$\alpha$ coupling coefficient $x_\alpha$ is directly proportional to the Ly$\alpha$ background flux $J_\alpha$, which we compute using:
\begin{equation}
J_\alpha(z) = \frac{c}{4 \pi} (1 + z)^3 \int_z^{z_\mathrm{max}} \mathrm{d}z' \, f_\alpha \, \Psi(z') \, \left|\frac{\mathrm{d}t'}{\mathrm{d}z'}\right|,
\end{equation}
where $f_\alpha$ is a normalization factor characterizing uncertainties in the properties of high-redshift stellar populations and radiative cascading effects. To determine the upper limit $z_\mathrm{max}$ of the integral, we assume that all continuum ionizing photons are absorbed by the IGM:
\begin{equation}
1 + z_\mathrm{max} = \frac{\nu_H}{\nu_\alpha} (1 + z),
\end{equation}
where $\nu_\alpha$ is the Ly$\alpha$ frequency and $\nu_H$ is the Lyman-limit frequency. The coupling coefficient is finally given by:
\begin{equation}
x_\alpha = 1.81 \times 10^{11} \frac{J_\alpha(z)}{\mathrm{cm}^{-2} \mathrm{s}^{-1} \mathrm{Hz}^{-1} \mathrm{sr}^{-1}}.
\end{equation}

Since the star formation rate density (SFRD), $\Psi(z)$, remains unconstrained at $z > 10$, we adopt a simple parametric form 
\begin{equation}
    \Psi = \Psi_0 e^{-\beta (z - z_0)},
\end{equation}
over the redshift range $z \sim 13-25$. Here, $\beta$ characterizes the steepness of the SFRD evolution, $z_0$ denotes the redshift at which X-ray heating begins, and $\Psi_0$ is the SFRD at $z_0$. We set $\Psi_0 = 5 \, \Psi_{\rm MD}(z_0)$, where $\Psi_{\rm MD}$ is the updated Madau–Dickinson SFRD \citep{Madau:2016jbv}, based on the original formulation of \citet{Madau:2014bja}.

\begin{figure}
    \centering
    \includegraphics[width=0.985\linewidth]{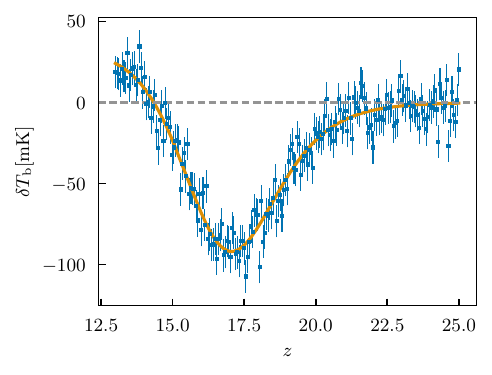}
    \vspace{-12pt}
    \caption{The 21-cm global signal for a fiducial SFRD of the form $\Psi = \Psi_0 e^{-\beta (z - z_0)}$. The redshift range is taken to be $z\sim13-25$. Other parameters are set to $\Psi_0 = 0.0066 \, M_{\odot} \rm  Mpc^{-3} yr^{-1}$, $\beta = 4/5$, $z_0 = 17$, $f_{\alpha} = 6 \times 10^{37}$, $f_{Xh} = 3 \times 10^{-2}$. A Gaussian instrumental noise of standard deviation $\Delta T_b = 10 \, \rm mK $ has been added to the signal. 
    }
    \label{fig: expected_21cm_sig}
\end{figure}

The Figure~\ref{fig: expected_21cm_sig} shows the modeled 21-cm global signal for the fiducial SFRD: $\Psi = \Psi_0 e^{-\beta (z - z_0)}$. 
The signal is computed assuming $\Psi_0 = 0.0066~M_{\odot}\mathrm{Mpc^{-3}yr^{-1}}$, $\beta = 4/5$, and $z_0 = 17$, together with $f_{\alpha} = 6 \times 10^{37}$ and $f_{Xh} = 3 \times 10^{-2}$. The SFRD parameters are motivated by observational and theoretical estimates from the excess abundance of high-redshift galaxies reported by JWST \citep{2024ApJ...960...56H,2024JCAP...07..078C,2026JCAP...01..008C}, while the adopted values of $f_{\alpha}$ and $f_{Xh}$ are guided by galaxy properties at low redshifts \citep{FurlanettoOhBriggs2006}. We assume Gaussian instrumental noise with a standard deviation of $\Delta T_b = 10\,\mathrm{mK}$, which is modestly better than that achieved by current experiments such as EDGES \citep{Bowman2018} and could be achieved through longer integration times. We assume that all astrophysical foregrounds and any potential instrumental systematics have been perfectly removed. Note that throughout this work, we use the \texttt{Planck18} Cosmology~\citep{Planck:2018vyg}.

Once we have modeled the 21-cm global signal, we infer the parameters generating this signal by adopting a Bayesian framework with a log-likelihood $\ln \mathcal{L}_{\rm 21-cm} \propto - \chi^2 / 2$, where
\begin{equation}
    \label{eq: 21cm_chi_sq}
    \chi^2 = \sum \left( \frac{\delta T_{\rm b} - \delta T_{\rm b, obs}}{\Delta T_{\rm b}} \right)^2.
\end{equation}
Here $\delta T_{\rm b, obs}$ denotes a synthetic (mock) observation of the 21-cm global signal generated using the model, $\Delta T_{\rm b}$ is the uncertainty in the measurement of $\delta T_{\rm b, obs}$, and $\delta T_{\rm b}$ is the modeled 21-cm signal. The posterior on the parameters is constructed via a simple application of Bayes' theorem.

\subsection{Simulating BBHs using Merger Rate}\label{subsec: sim_bbhs}
Let $dN$ be the total number of BBH mergers that happened between redshifts $z_m$ and $z_m + dz_m$ and observation time $t_{\rm o}$ and $t_{\rm o} + dt_{\rm o}$. The merger rate of these events is then given by $R (z_m) \equiv dN/[dt_{\rm o} dz_m]$. Let $dt_{\rm s}$ be the corresponding duration in the source frame and $dV_{\rm c}/dz$ be the differential comoving volume. Then we can write~\citep{Vitale:2018yhm, Fishbach:2018edt, KAGRA:2021duu}
\begin{equation}
    \label{eq: df_mr}
    R(z_m) = \frac{1}{1 + z_m}\frac{dV_{\rm c}}{dz}(z_m) \mathcal{R}(z_m)\,,
\end{equation}
where  $dt_{\rm o} = dt_{\rm s} (1 + z_m)$ and $\mathcal{R}(z_m) \equiv dN / [dV_{\rm c}dt_{\rm s}]$ is the source-frame volumetric merger rate, which is given by
\begin{equation}
    \label{eq: sf_vol_mr}
    \mathcal{R} (z_m) = \int_{z_m}^{\infty} dz \frac{dt_f}{dz} p(t_d) \mathcal{R}_f (z)\,,
\end{equation}
where $t_d$ is the delay time $t_f(z) - t(z_m)$, i.e., the time elapsed between the formation and merger of a BBH, and $p(t_d)$ and $\mathcal{R}_f (z) \equiv dN_f/[dV_{\rm c}dt_f]$ are the delay time distribution and volumetric formation rate of BBHs, respectively. $\mathcal{R}_f (z)$ depends\footnote{Apart from these quantities, $\mathcal{R}_f$ also depends on the binary fraction and binary formation channel.} on the star formation rate density $\Psi(z)$, initial mass function (IMF) $\xi$, and BBH formation efficiency $f_{\rm BBH} (z)$ and hence
\begin{equation}
    \label{eq: bbh_vol_fr}
    \mathcal{R}_f(z) \propto  \xi f_{\rm BBH} (z) \Psi(z).
\end{equation}

Following \citep{Vitale:2018yhm}, we assume a fixed $16-15 \, M_{\odot}$ BBH delta-function like mass distribution. As shown in \citep{Hall:2022dik}, this corresponds to a detectable horizon for CE that far exceeds the Cosmic Dawn and Dark Ages epochs, thus ensuring that selection effects can be safely neglected. Redshifts are drawn from Equation~\ref{eq: df_mr}, assuming $f_{\rm BBH} = 1$ for all $z$. Since the delay-time distribution of BBHs has fairly weak constraints \citep{Vijaykumar:2023bgs}, we fix a delay time of $1\,\rm Gyr$, i.e., $p(t_d) = \delta_D(t_d - 1 \,\rm Gyr)$. We fix the proportionality constant of Equation~\ref{eq: bbh_vol_fr} such that we get $\mathcal{R}(z_m = 4.822) = 20 \, \rm Gpc^{-3}yr^{-1}$  --- this ensures that $\mathcal{O}$(100) events will be observed from $z = 13 - 25$ in a $\sim 70$-days observation period, and $\mathcal{O}(1000)$ in a $\sim 2$-year period.

\subsection{Hierarchical Inference Likelihood} \label{subsec: hyp_like}
We can write the hyper-likelihood for the $i$th BBH merger at redshift $z_i$ as
\begin{equation}
    \label{eq: sing_ev_hlike_1}
    \mathcal{L}(d_i \vert \{ \Psi_0, \beta \})  = \int dz'_m \mathcal{L}(d_i \vert z'_m) R(z'_m \vert \{\Psi_0, \beta\})
\end{equation}
where $\mathcal{L}(d_i \vert z'_m)$ is the likelihood of getting the data for the $i$th merger, which we approximate as a Gaussian using the Fisher matrix~\citep{1994PhRvD..49.2658C}. 
Since we do not get the redshift $z_m$ of an event directly from the GW signal, we write $\mathcal{L}(d_i \vert z'_m)$ in terms of $\mathcal{L}(d_i \vert D_{{\rm L}}(z'_m))$ as:
\begin{equation}
    \label{ep: z_like_from_dl}
    \mathcal{L}(d_i \vert z'_m) = \mathcal{L}(d_i \vert D_{{\rm L}}(z'_m)) \propto \exp\left\{- \frac{(D'_{\rm L} (z'_m) - D_{{\rm L},i})^2}{2 \sigma_{D_{{\rm L},i}}} \right\} 
\end{equation}
where $D_{{\rm L},i}$ is the luminosity distance of the $i$th event and $\sigma_{D_{{\rm L},i}}$ is the projected error in its measurement obtained from the Fisher matrix.

As the observation of any BBH merger is an event independent of other mergers, the total hyper-likelihood will just be a product of individual hyper-likelihoods:
\begin{equation}
\mathcal{L}_{\rm tot}(\boldsymbol{d} \vert \{\Psi_0, \beta\}) = \prod_i^N \mathcal{L}(d_i \vert \{ \Psi_0, \beta \}),
\end{equation}
where $N$ is the number of total BBH mergers. Therefore, we can write the total log-hyper-likelihood of all events as:
\begin{equation}
    \ln \mathcal{L}_{\rm tot}(\boldsymbol{d} \vert \{\Psi_0, \beta\}) \propto \sum_i^N \ln \mathcal{L}(d_i \vert \{ \Psi_0, \beta \})
\end{equation}
In general, the total hyper-posterior is given by (see Appendix~\ref{app: hier} for more details):
\begin{equation}
    p_{\rm tot}(\{\Psi_0, \beta\} \vert \boldsymbol{d}) \propto \mathcal{L}_{\rm tot}(\boldsymbol{d} \vert \{\Psi_0, \beta\}) \pi(\Psi_0)\pi(\beta)
\end{equation}
where $\pi(\ldots)$ are the priors.
However, the observation of BBHs is a Poisson process. Therefore:
\begin{equation}
    \label{eq: hyper_post}
    p_{\rm tot}(\{\Psi_0, \beta\} \vert \boldsymbol{d}) \propto \zeta^N e^{-\zeta} \mathcal{L}_{\rm tot}(\boldsymbol{d} \vert \{\Psi_0, \beta\}) \pi(\Psi_0)\pi(\beta),
\end{equation}
where $\zeta = T_{\rm obs} \int dz'_m R (z'_m \vert \{\Psi_0, \beta\})$ is the expected number of events for a given model and $T_{\rm obs}$ is the time for observing $N$ events. In the absence of selection effects, $\zeta^N$ cancels out with the normalization of $R(z'_m \vert \{\Psi_0, \beta \})$. Taking this into consideration, the final expression of total hyper-likelihood simplifies to:
\begin{equation}
    \label{eq: final_tot_llog_hlike}
    \ln \mathcal{L}_{\rm tot}(\boldsymbol{d} \vert \{\Psi_0, \beta\}) \propto -\zeta + \sum_i^N \ln \mathcal{L}(d_i \vert \{ \Psi_0, \beta \}) 
\end{equation}

\begin{figure}
    \centering
    \includegraphics[width=0.985\linewidth]{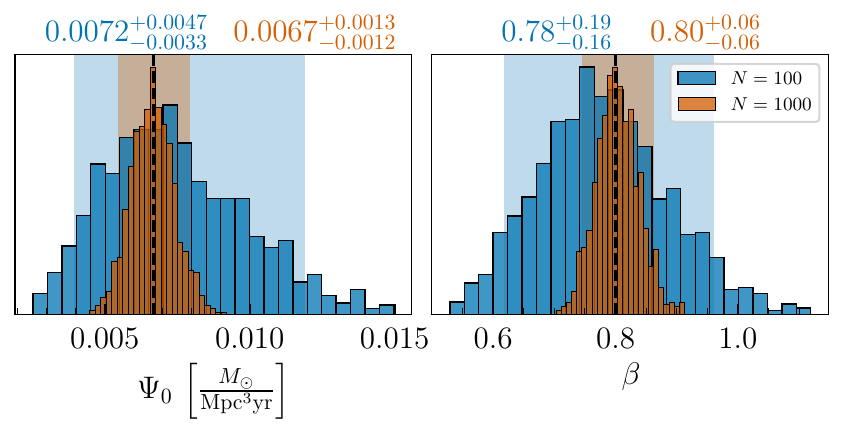}
    \vspace{-12pt}
    \caption{The $1d$-posteriors of the SFRD parameters inferred exclusively by GW hierarchical inference on 100 (blue) and 1000 (brown) BBH events. The black dashed lines represent the true values, while the light blue and brown shaded regions represent the 90\% credible intervals.}
    \label{fig: gw_post}
\end{figure}

\section{Constraining the SFRD and Inferring the 21-cm global signal parameters}\label{subsec: const_sfrd}
Let $\mathcal{M}$, $\eta$, and $D_{\rm L}$ be the redshifted chirp mass, symmetric mass ratio, and luminosity distance of a BBH, respectively. We perform the Fisher matrix analysis~\citep{1994PhRvD..49.2658C} on the parameter space $\Theta=\{{\ln \mathcal{M}, \, \ln \eta,\, \ln D_{\rm L}}\}$ using the \textsc{IMRPhenomD} waveform~\citep{Husa:2015iqa, Khan:2015jqa} and assuming the CE band\footnote{We use the compact-binary optimized 40 km baseline design~\citep{Srivastava:2022slt} PSD of CE available at \href{https://dcc.cosmicexplorer.org/CE-T2000017/public}{URL}.}. 
The inverse of the Fisher matrix gives us the Covariance matrix, which contains the errors in the measurement of these parameters (see Appendix~\ref{app: fisher} for more details). 
We then use these projected errors in $D_{\rm L}$ in the Equation~\ref{eq: final_tot_llog_hlike} and get the constraints on $\Psi_0$ and $\beta$.

We use these constraints of the SFRD parameters as priors while inferring the parameters generating the 21-cm global signal with log-likelihood $\ln \mathcal{L}_{\rm 21-cm} \propto - \chi^2 / 2$, where $\chi^2$ is given by Equation~\ref{eq: 21cm_chi_sq}. Specifically, we use a multivariate normal distribution $\mathcal{N}(\Psi_0, \beta)$ as prior on $\Psi_0$ and $\beta$ constructed using the posteriors obtained from GW population inference; and uniform priors $\mathcal{U}(35, \, 40)$ and $\mathcal{U}(-5,\, 0)$ on $\log f_{\alpha}$ and $\log f_{Xh}$, respectively.

\section{Results}\label{sec: results}
We simulated two populations of BBHs following the prescription in section~\ref{subsec: sim_bbhs} and performed the Fisher matrix analysis as described in section~\ref{subsec: const_sfrd}, assuming a 3 CE detector configuration to get the projected errors in the measurement of $D_{\rm L}$. We then performed the hierarchical inference to obtain the posteriors of $\Psi_0$ and $\beta$. Figure~\ref{fig: gw_post} shows the inferred posteriors of $\Psi_0$ and $\beta$ for the populations of 100 and 1000 GW events -- we have varied the number of events in the population and found that $\sim 90$ events are enough to constrain both SFRD parameters simultaneously (see Figure~\ref{fig: varying_nev} of the Appendix~\ref{app: additional_plots}). 
We summarize the constraints in Table~\ref{tab: gw_res}. Specifically, we find that $\Psi_0$ and $\beta$ both are measured at a 90\% credible level, and constraints become tighter after increasing the number of events in the population. 

\begin{table}
\centering
\begin{tabular}{ccc}
\hline
$N$ 
& $\Psi_0 \,[M_{\odot} \rm Mpc^{-3} yr^{-1}]$ 
& $\beta$ \\
\hline

100 
& $0.0072^{+0.0047}_{-0.0033}$ 
& $0.78^{+0.19}_{-0.16}$ \\
\hline

1000 
& $0.0067^{+0.0013}_{-0.0012}$ 
& $0.80^{+0.06}_{-0.06}$ \\
\hline

\end{tabular}
\caption{
Projected constraints on the SFRD parameters from hierarchical inference on the population of simulated GW events (see Figure~\ref{fig: gw_post}).
}
\label{tab: gw_res}
\end{table}

\begin{figure*}
    \centering
    \includegraphics[width=0.985\linewidth]{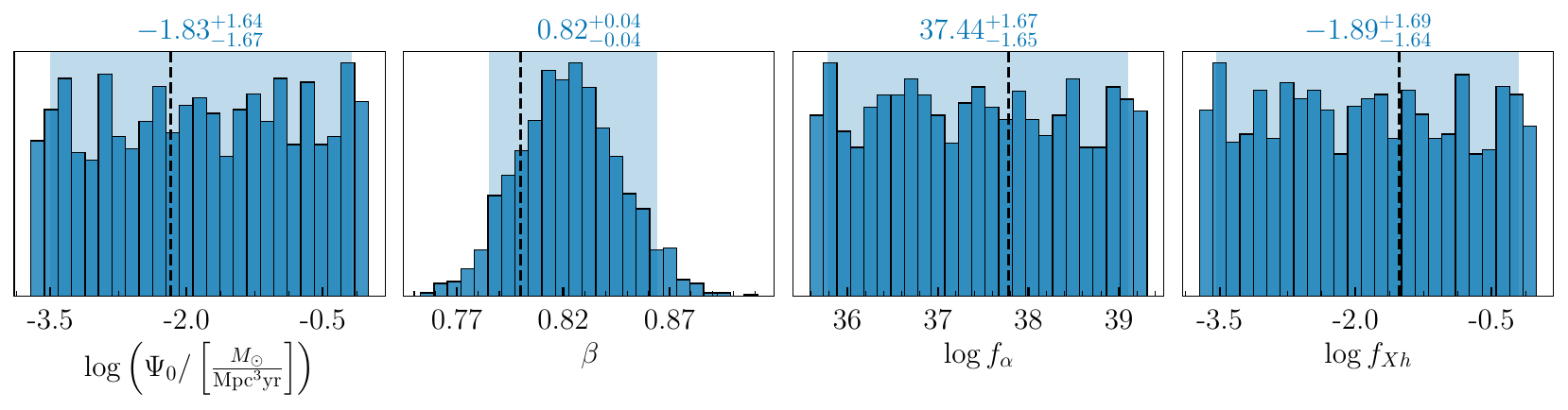}
    \includegraphics[width=0.985\linewidth]{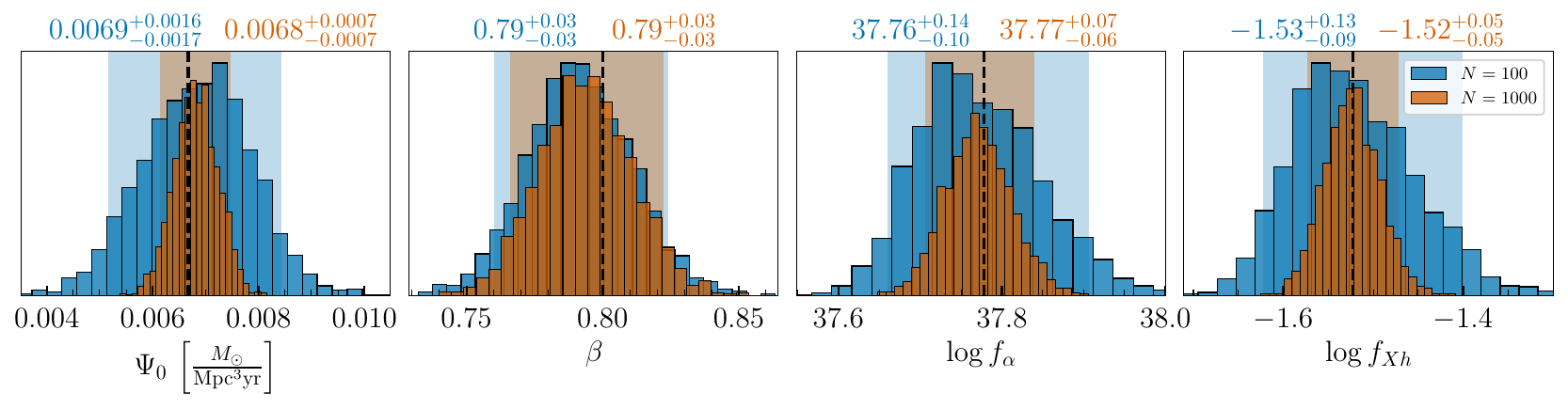}
    \vspace{-12pt}
    \caption{{\it Top panel}: The $1d$-posteriors of the 21-cm global signal parameters using only the data from the 21cm observation, with no complementary GW information. {\it Bottom panel}: The inferred posteriors of 21-cm global signal parameters obtained using both the 21-cm signal observation, as well as the GW hierarchical inference results shown in Figure~\ref{fig: gw_post}. Blue histograms correspond to inference involving 100 GW events, while the brown histograms correspond to 1000 GW events. The black dashed line and the light shaded regions have the same meaning as in Figure~\ref{fig: gw_post}.}
    \label{fig: 21cm_post}
\end{figure*}

The {\it top panel} of Figure~\ref{fig: 21cm_post} shows the inferred posteriors of the 21-cm global signal parameters. It shows that none of the parameters except $\beta$ are measured, even at a 90\% credible level. Indeed, there is little difference between posterior and prior for these parameters. The parameters $\Psi_0$, $f_\alpha$, and $f_{Xh}$ primarily act as normalization factors for the Lyman-$\alpha$ coupling and X-ray heating rates. Consequently, they suffer from severe amplitude degeneracies among themselves and cannot be constrained independently using the 21-cm signal alone. In contrast, the parameter $\beta$ characterizes the steepness of the SFRD evolution and therefore controls the distinct shape and width of the 21-cm absorption trough. This shape information cannot be mimicked by the other efficiency parameters, allowing $\beta$ to remain measurable even when the overall amplitude is unconstrained.

On the other hand, when we use the GW posteriors on $\Psi_0$ and $\beta$  as priors while inferring the other 21-cm global signal parameters, the inference on these parameters improves significantly, as shown in the {\it bottom panel} of  Figure~\ref{fig: 21cm_post}. We summarize the parameter inferences in Table~\ref{tab: 21cm_res} and show a complementary corner plot of the posteriors for the 100 events scenario in Appendix~\ref{app: additional_plots}. We find drastic improvements in the constraints of 21-cm global signal parameters $\Psi_0$, $\beta$, $\log f_{\alpha}$, and $\log f_{Xh}$ compared to {\it top panel} of Figure~\ref{fig: 21cm_post} which pertained to the inference of these parameters using only the 21-cm signal. Moreover, we also find that the constraints on $\Psi_0$ and $\beta$ become even more stringent than those obtained from the GW-observed BBHs. Increasing the number of BBH events from $100$ to $1000$ tightens all the parameter posteriors, except $\beta$, which was already well constrained from the 21-cm signal. Because the 21-cm signal alone has placed a much tighter constraint on the steepness parameter $\beta$ than the same from GW hierarchical inference\footnote{This statement holds only for the number of events considered in this analysis; constraints from GW hierarchical inference would become increasingly stringent as $N$ grows because the errors scale as $\sim 1/\sqrt{N}$.} (see Table~\ref{tab: gw_res}), the combined posterior for $\beta$ remain entirely dominated by the 21-cm data.

\begin{table}
\centering
\begin{tabular}{ccccc}
\hline
$N$ 
& $\Psi_0 \,[M_{\odot} \rm Mpc^{-3} yr^{-1}]$ 
& $\beta$ 
& $\log f_{\alpha}$
& $\log f_{Xh}$  \\
\hline

100 
& $0.0069^{+0.0016}_{-0.0017}$ 
& $0.79^{+0.03}_{-0.03}$ 
& $37.76^{+0.14}_{-0.10}$ 
& $-1.53^{+0.13}_{-0.09}$ \\
\hline

1000 
& $0.0068^{+0.0007}_{-0.0007}$ 
& $0.79^{+0.03}_{-0.03}$ 
& $37.77^{+0.07}_{-0.06}$
& $-1.52^{+0.05}_{-0.05}$ \\
\hline

\end{tabular}
\caption{
Projected constraints on the 21-cm global signal parameters (see the {\it bottom panel} of Figure~\ref{fig: 21cm_post}).
}
\label{tab: 21cm_res}
\end{table}

\section{Discussion}

Measuring the star-formation rate of the early universe post-recombination is essential to accurately constrain the parameters of the reionization history. GWs from mergers of binaries that are formed during the earliest epochs of galaxy formation and merge at high redshift will be accessible in next-generation GW experiments, giving us an independent probe of the star-formation history. Our work shows that even with $100$ events at high redshifts, one can begin to break degeneracies in the reionization model for the global 21-cm signal. 

In this {\it Letter}, we demonstrate that combining third-generation GW observations of binary black holes (BBHs) with measurements of the 21-cm signal can substantially improve inference of 21-cm signal parameters. In particular, we constrain the star-formation rate density (SFRD) parameters using GW observations from Cosmic Explorer (CE) by performing hierarchical inference on a simulated population of BBHs. As a proof of concept, the population is sampled only in redshift (z $\sim 13\!-\!25$), while the BBH component masses and the delay time are fixed to values optimized for CE sensitivity, following \citet{Vitale:2018yhm}. These simplifying assumptions are adopted to reduce the computational cost associated with high-dimensional population inference. They are also well justified given that this work represents the first demonstration of the potential gains from synergizing observations of resolvable high-redshift BBHs with those of the 21-cm signal, and given that the intrinsic BBH population properties and delay-time distribution—particularly at high redshift—are currently essentially unconstrained.

In future work, we will extend this framework by adopting physically motivated distributions for BBH delay times and BBH masses at high redshift. In a realistic detection scenario, the observed BBH population will comprise multiple formation channels and redshift origins \citep{Zevin:2020gbd}, with only a subset of detected systems probing the Dark Ages and Cosmic Dawn. Accurately modeling such a population will therefore require a parameterized treatment that accounts for multiple provenances and their associated branching fractions, together with corresponding mass, spin, and delay-time distributions. Incorporating this additional realism and population complexity is a natural next step and will be pursued in forthcoming work.

\section*{Acknowledgements}

We thank organizers and participants of the ``Pune-Mumbai Cosmology and Astroparticle Physics (PMCAP) Meeting'' held at IISER Pune, during which this project was initiated. SJK acknowledges support from ANRF/SERB Grants SRG/2023/000419 and MTR/2023/000086. SA is supported by ANRF/SERB Grants SRG/2023/001563. MKS thanks the Science and Technology Facilities Council (STFC) for funding through the grant ST/Y004272/1.

\section*{Data Availability}

 The data pertaining to this paper will be made available upon reasonable request.

\section*{Software} 
\textsc{Bilby}~\citep{bilby_paper}, \textsc{lalsuite}~\citep{lalsuite},
\textsc{NumPy} \citep{vanderWalt:2011bqk}, \textsc{SciPy} \citep{Virtanen:2019joe}, \textsc{astropy} \citep{2013A&A...558A..33A, 2018AJ....156..123A}, \textsc{Matplotlib} \citep{Hunter:2007}, \textsc{seaborn}~\citep{Waskom2021}, and \textsc{jupyter} \citep{jupyter}.



\bibliographystyle{mnras}
\bibliography{refer} 

\begin{thebibliography}{}
\makeatletter
\relax
\def\mn@urlcharsother{\let\do\@makeother \do\$\do\&\do\#\do\^\do\_\do\%\do\~}
\def\mn@doi{\begingroup\mn@urlcharsother \@ifnextchar [ {\mn@doi@}
  {\mn@doi@[]}}
\def\mn@doi@[#1]#2{\def\@tempa{#1}\ifx\@tempa\@empty \href
  {http://dx.doi.org/#2} {doi:#2}\else \href {http://dx.doi.org/#2} {#1}\fi
  \endgroup}
\def\mn@eprint#1#2{\mn@eprint@#1:#2::\@nil}
\def\mn@eprint@arXiv#1{\href {http://arxiv.org/abs/#1} {{\tt arXiv:#1}}}
\def\mn@eprint@dblp#1{\href {http://dblp.uni-trier.de/rec/bibtex/#1.xml}
  {dblp:#1}}
\def\mn@eprint@#1:#2:#3:#4\@nil{\def\@tempa {#1}\def\@tempb {#2}\def\@tempc
  {#3}\ifx \@tempc \@empty \let \@tempc \@tempb \let \@tempb \@tempa \fi \ifx
  \@tempb \@empty \def\@tempb {arXiv}\fi \@ifundefined
  {mn@eprint@\@tempb}{\@tempb:\@tempc}{\expandafter \expandafter \csname
  mn@eprint@\@tempb\endcsname \expandafter{\@tempc}}}

\bibitem[\protect\citeauthoryear{Abbott et~al.}{Abbott
  et~al.}{2023}]{KAGRA:2021duu}
Abbott R.,  et~al., 2023, \mn@doi [Phys. Rev. X] {10.1103/PhysRevX.13.011048},
  13, 011048

\bibitem[\protect\citeauthoryear{{Adamo} et~al.,}{{Adamo}
  et~al.}{2025}]{2025NatAs...9.1134A}
{Adamo} A.,  et~al., 2025, \mn@doi [Nature Astronomy]
  {10.1038/s41550-025-02624-5}, \href
  {https://ui.adsabs.harvard.edu/abs/2025NatAs...9.1134A} {9, 1134}

\bibitem[\protect\citeauthoryear{Aghanim et~al.}{Aghanim
  et~al.}{2020}]{Planck:2018vyg}
Aghanim N.,  et~al., 2020, \mn@doi [Astron. Astrophys.]
  {10.1051/0004-6361/201833910}, 641, A6

\bibitem[\protect\citeauthoryear{Ashton et~al.}{Ashton
  et~al.}{2019}]{bilby_paper}
Ashton G.,  et~al., 2019, \mn@doi [Astrophys. J. Suppl.]
  {10.3847/1538-4365/ab06fc}, 241, 27

\bibitem[\protect\citeauthoryear{{Barkana} \& {Loeb}}{{Barkana} \&
  {Loeb}}{2001}]{2001PhR...349..125B}
{Barkana} R.,  {Loeb} A.,  2001, \mn@doi [\physrep]
  {10.1016/S0370-1573(01)00019-9}, \href
  {https://ui.adsabs.harvard.edu/abs/2001PhR...349..125B} {349, 125}

\bibitem[\protect\citeauthoryear{{Bowman}, {Rogers}  \& {Hewitt}}{{Bowman}
  et~al.}{2008}]{2008ApJ...676....1B}
{Bowman} J.~D.,  {Rogers} A. E.~E.,   {Hewitt} J.~N.,  2008, \mn@doi [\apj]
  {10.1086/528675}, \href
  {https://ui.adsabs.harvard.edu/abs/2008ApJ...676....1B} {676, 1}

\bibitem[\protect\citeauthoryear{{Bowman}, {Rogers}, {Monsalve}, {Mozdzen}  \&
  {Mahesh}}{{Bowman} et~al.}{2018}]{Bowman2018}
{Bowman} J.~D.,  {Rogers} A. E.~E.,  {Monsalve} R.~A.,  {Mozdzen} T.~J.,
  {Mahesh} N.,  2018, \mn@doi [\nat] {10.1038/nature25792}, \href
  {https://ui.adsabs.harvard.edu/abs/2018Natur.555...67B} {555, 67}

\bibitem[\protect\citeauthoryear{{Chakraborty} \& {Choudhury}}{{Chakraborty} \&
  {Choudhury}}{2024}]{2024JCAP...07..078C}
{Chakraborty} A.,  {Choudhury} T.~R.,  2024, \mn@doi [\jcap]
  {10.1088/1475-7516/2024/07/078}, \href
  {https://ui.adsabs.harvard.edu/abs/2024JCAP...07..078C} {2024, 078}

\bibitem[\protect\citeauthoryear{{Chakraborty} \& {Choudhury}}{{Chakraborty} \&
  {Choudhury}}{2026}]{2026JCAP...01..008C}
{Chakraborty} A.,  {Choudhury} T.~R.,  2026, \mn@doi [\jcap]
  {10.1088/1475-7516/2026/01/008}, \href
  {https://ui.adsabs.harvard.edu/abs/2026JCAP...01..008C} {2026, 008}

\bibitem[\protect\citeauthoryear{Chatterjee \& Choudhury}{Chatterjee \&
  Choudhury}{2023}]{ChatterjeeChoudhury2023}
Chatterjee A.,  Choudhury T.~R.,  2023, \mn@doi [Mon. Not. Roy. Astron. Soc.]
  {10.1093/mnras/stad3930}, 527, 10777

\bibitem[\protect\citeauthoryear{{Chatterjee}, {Dayal}, {Choudhury}  \&
  {Hutter}}{{Chatterjee} et~al.}{2019}]{ChatterjeeDayalChoudhuryHutter2019}
{Chatterjee} A.,  {Dayal} P.,  {Choudhury} T.~R.,   {Hutter} A.,  2019, \mn@doi
  [\mnras] {10.1093/mnras/stz1444}, \href
  {https://ui.adsabs.harvard.edu/abs/2019MNRAS.487.3560C} {487, 3560}

\bibitem[\protect\citeauthoryear{{Chatterjee}, {Dayal}, {Choudhury}  \&
  {Schneider}}{{Chatterjee}
  et~al.}{2020}]{ChatterjeeDayalChoudhurySchneider2020}
{Chatterjee} A.,  {Dayal} P.,  {Choudhury} T.~R.,   {Schneider} R.,  2020,
  \mn@doi [\mnras] {10.1093/mnras/staa1609}, \href
  {https://ui.adsabs.harvard.edu/abs/2020MNRAS.496.1445C} {496, 1445}

\bibitem[\protect\citeauthoryear{{Choudhury}}{{Choudhury}}{2022}]{2022GReGr..54..102C}
{Choudhury} T.~R.,  2022, \mn@doi [General Relativity and Gravitation]
  {10.1007/s10714-022-02987-4}, \href
  {https://ui.adsabs.harvard.edu/abs/2022GReGr..54..102C} {54, 102}

\bibitem[\protect\citeauthoryear{{Cutler} \& {Flanagan}}{{Cutler} \&
  {Flanagan}}{1994}]{1994PhRvD..49.2658C}
{Cutler} C.,  {Flanagan} {\'E}.~E.,  1994, \mn@doi [\prd]
  {10.1103/PhysRevD.49.2658}, \href
  {https://ui.adsabs.harvard.edu/abs/1994PhRvD..49.2658C} {49, 2658}

\bibitem[\protect\citeauthoryear{{Dhandha} et~al.,}{{Dhandha}
  et~al.}{2025}]{2025MNRAS.542.2292D}
{Dhandha} J.,  et~al., 2025, \mn@doi [\mnras] {10.1093/mnras/staf1359}, \href
  {https://ui.adsabs.harvard.edu/abs/2025MNRAS.542.2292D} {542, 2292}

\bibitem[\protect\citeauthoryear{{Field}}{{Field}}{1958}]{1958PIRE...46..240F}
{Field} G.~B.,  1958, \mn@doi [Proceedings of the IRE]
  {10.1109/JRPROC.1958.286741}, \href
  {https://ui.adsabs.harvard.edu/abs/1958PIRE...46..240F} {46, 240}

\bibitem[\protect\citeauthoryear{Fishbach, Holz  \& Farr}{Fishbach
  et~al.}{2018}]{Fishbach:2018edt}
Fishbach M.,  Holz D.~E.,   Farr W.~M.,  2018, \mn@doi [Astrophys. J. Lett.]
  {10.3847/2041-8213/aad800}, 863, L41

\bibitem[\protect\citeauthoryear{Furlanetto, Oh  \& Briggs}{Furlanetto
  et~al.}{2006}]{FurlanettoOhBriggs2006}
Furlanetto S.,  Oh S.~P.,   Briggs F.,  2006, \mn@doi [Phys. Rept.]
  {10.1016/j.physrep.2006.08.002}, 433, 181

\bibitem[\protect\citeauthoryear{Hall}{Hall}{2022}]{Hall:2022dik}
Hall E.~D.,  2022, \mn@doi [Galaxies] {10.3390/galaxies10040090}, 10, 90

\bibitem[\protect\citeauthoryear{{Harikane}, {Nakajima}, {Ouchi}, {Umeda},
  {Isobe}, {Ono}, {Xu}  \& {Zhang}}{{Harikane}
  et~al.}{2024}]{2024ApJ...960...56H}
{Harikane} Y.,  {Nakajima} K.,  {Ouchi} M.,  {Umeda} H.,  {Isobe} Y.,  {Ono}
  Y.,  {Xu} Y.,   {Zhang} Y.,  2024, \mn@doi [\apj] {10.3847/1538-4357/ad0b7e},
  \href {https://ui.adsabs.harvard.edu/abs/2024ApJ...960...56H} {960, 56}

\bibitem[\protect\citeauthoryear{Hild et~al.}{Hild et~al.}{2011}]{Hild:2010id}
Hild S.,  et~al., 2011, \mn@doi [Class. Quant. Grav.]
  {10.1088/0264-9381/28/9/094013}, 28, 094013

\bibitem[\protect\citeauthoryear{Hunter}{Hunter}{2007}]{Hunter:2007}
Hunter J.~D.,  2007, \mn@doi [Computing in Science \& Engineering]
  {10.1109/MCSE.2007.55}, 9, 90

\bibitem[\protect\citeauthoryear{Husa, Khan, Hannam, P{\"u}rrer, Ohme,
  Jim{\'e}nez~Forteza  \& Boh{\'e}}{Husa et~al.}{2016}]{Husa:2015iqa}
Husa S.,  Khan S.,  Hannam M.,  P{\"u}rrer M.,  Ohme F.,  Jim{\'e}nez~Forteza
  X.,   Boh{\'e} A.,  2016, \mn@doi [Phys. Rev. D]
  {10.1103/PhysRevD.93.044006}, 93, 044006

\bibitem[\protect\citeauthoryear{Khan, Husa, Hannam, Ohme, P{\"u}rrer,
  Jim{\'e}nez~Forteza  \& Boh{\'e}}{Khan et~al.}{2016}]{Khan:2015jqa}
Khan S.,  Husa S.,  Hannam M.,  Ohme F.,  P{\"u}rrer M.,  Jim{\'e}nez~Forteza
  X.,   Boh{\'e} A.,  2016, \mn@doi [Phys. Rev. D]
  {10.1103/PhysRevD.93.044007}, 93, 044007

\bibitem[\protect\citeauthoryear{Kluyver et~al.,}{Kluyver
  et~al.}{2016}]{jupyter}
Kluyver T.,  et~al., 2016, in Loizides F.,  Scmidt B.,  eds, Positioning and
  Power in Academic Publishing: Players, Agents and Agendas. IOS Press,
  Netherlands, pp 87--90, \url {https://eprints.soton.ac.uk/403913/}

\bibitem[\protect\citeauthoryear{{LIGO Scientific Collaboration}, {Virgo
  Collaboration}  \& {KAGRA Collaboration}}{{LIGO Scientific Collaboration}
  et~al.}{2018}]{lalsuite}
{LIGO Scientific Collaboration} {Virgo Collaboration}  {KAGRA Collaboration}
  2018, {LVK} {A}lgorithm {L}ibrary - {LALS}uite, Free software (GPL),
  \mn@doi{10.7935/GT1W-FZ16}

\bibitem[\protect\citeauthoryear{Loredo}{Loredo}{2004}]{Loredo:2004nn}
Loredo T.~J.,  2004, \mn@doi [AIP Conf. Proc.] {10.1063/1.1835214}, 735, 195

\bibitem[\protect\citeauthoryear{Madau \& Dickinson}{Madau \&
  Dickinson}{2014}]{Madau:2014bja}
Madau P.,  Dickinson M.,  2014, \mn@doi [Ann. Rev. Astron. Astrophys.]
  {10.1146/annurev-astro-081811-125615}, 52, 415

\bibitem[\protect\citeauthoryear{Madau \& Fragos}{Madau \&
  Fragos}{2017}]{Madau:2016jbv}
Madau P.,  Fragos T.,  2017, \mn@doi [Astrophys. J.]
  {10.3847/1538-4357/aa6af9}, 840, 39

\bibitem[\protect\citeauthoryear{{Madau}, {Meiksin}  \& {Rees}}{{Madau}
  et~al.}{1997}]{1997ApJ...475..429M}
{Madau} P.,  {Meiksin} A.,   {Rees} M.~J.,  1997, \mn@doi [\apj]
  {10.1086/303549}, \href
  {https://ui.adsabs.harvard.edu/abs/1997ApJ...475..429M} {475, 429}

\bibitem[\protect\citeauthoryear{{Mittal} \& {Kulkarni}}{{Mittal} \&
  {Kulkarni}}{2022}]{2022MNRAS.515.2901M}
{Mittal} S.,  {Kulkarni} G.,  2022, \mn@doi [\mnras] {10.1093/mnras/stac1961},
  \href {https://ui.adsabs.harvard.edu/abs/2022MNRAS.515.2901M} {515, 2901}

\bibitem[\protect\citeauthoryear{Price-Whelan et~al.}{Price-Whelan
  et~al.}{2018}]{2018AJ....156..123A}
Price-Whelan A.~M.,  et~al., 2018, \mn@doi [Astron. J.]
  {10.3847/1538-3881/aabc4f}, 156, 123

\bibitem[\protect\citeauthoryear{{Price} et~al.,}{{Price}
  et~al.}{2018}]{Price2018}
{Price} D.~C.,  et~al., 2018, \mn@doi [\mnras] {10.1093/mnras/sty1244}, \href
  {https://ui.adsabs.harvard.edu/abs/2018MNRAS.478.4193P} {478, 4193}

\bibitem[\protect\citeauthoryear{{Pritchard} \& {Loeb}}{{Pritchard} \&
  {Loeb}}{2012}]{2012RPPh...75h6901P}
{Pritchard} J.~R.,  {Loeb} A.,  2012, \mn@doi [Reports on Progress in Physics]
  {10.1088/0034-4885/75/8/086901}, \href
  {https://ui.adsabs.harvard.edu/abs/2012RPPh...75h6901P} {75, 086901}

\bibitem[\protect\citeauthoryear{Reitze et~al.}{Reitze
  et~al.}{2019}]{Reitze:2019iox}
Reitze D.,  et~al., 2019, Bull. Am. Astron. Soc., 51, 035

\bibitem[\protect\citeauthoryear{Robitaille et~al.}{Robitaille
  et~al.}{2013}]{2013A&A...558A..33A}
Robitaille T.~P.,  et~al., 2013, \mn@doi [Astron. Astrophys.]
  {10.1051/0004-6361/201322068}, 558, A33

\bibitem[\protect\citeauthoryear{{Singh} et~al.,}{{Singh}
  et~al.}{2022}]{ssingh2022}
{Singh} S.,  et~al., 2022, \mn@doi [Nature Astronomy]
  {10.1038/s41550-022-01610-5}, \href
  {https://ui.adsabs.harvard.edu/abs/2022NatAs...6..607S} {6, 607}

\bibitem[\protect\citeauthoryear{Srivastava et~al.,}{Srivastava
  et~al.}{2022}]{Srivastava:2022slt}
Srivastava V.,  et~al., 2022, \mn@doi [Astrophys. J.]
  {10.3847/1538-4357/ac5f04}, 931, 22

\bibitem[\protect\citeauthoryear{Talbot \& Thrane}{Talbot \&
  Thrane}{2017}]{Talbot:2017yur}
Talbot C.,  Thrane E.,  2017, \mn@doi [Phys. Rev. D]
  {10.1103/PhysRevD.96.023012}, 96, 023012

\bibitem[\protect\citeauthoryear{Thrane \& Talbot}{Thrane \&
  Talbot}{2019}]{Thrane:2018qnx}
Thrane E.,  Talbot C.,  2019, \mn@doi [Publ. Astron. Soc. Austral.]
  {10.1017/pasa.2019.2}, 36, e010

\bibitem[\protect\citeauthoryear{Vijaykumar, Fishbach, Adhikari  \&
  Holz}{Vijaykumar et~al.}{2024}]{Vijaykumar:2023bgs}
Vijaykumar A.,  Fishbach M.,  Adhikari S.,   Holz D.~E.,  2024, \mn@doi
  [Astrophys. J.] {10.3847/1538-4357/ad6140}, 972, 157

\bibitem[\protect\citeauthoryear{Virtanen et~al.}{Virtanen
  et~al.}{2020}]{Virtanen:2019joe}
Virtanen P.,  et~al., 2020, \mn@doi [Nature Meth.] {10.1038/s41592-019-0686-2}

\bibitem[\protect\citeauthoryear{Vitale, Farr, Ng  \& Rodriguez}{Vitale
  et~al.}{2019}]{Vitale:2018yhm}
Vitale S.,  Farr W.~M.,  Ng K.,   Rodriguez C.~L.,  2019, \mn@doi [Astrophys.
  J. Lett.] {10.3847/2041-8213/ab50c0}, 886, L1

\bibitem[\protect\citeauthoryear{Waskom}{Waskom}{2021}]{Waskom2021}
Waskom M.~L.,  2021, \mn@doi [Journal of Open Source Software]
  {10.21105/joss.03021}, 6, 3021

\bibitem[\protect\citeauthoryear{{Wouthuysen}}{{Wouthuysen}}{1952}]{1952AJ.....57R..31W}
{Wouthuysen} S.~A.,  1952, \mn@doi [\aj] {10.1086/106661}, \href
  {https://ui.adsabs.harvard.edu/abs/1952AJ.....57R..31W} {57, 31}

\bibitem[\protect\citeauthoryear{Zevin et~al.,}{Zevin
  et~al.}{2021}]{Zevin:2020gbd}
Zevin M.,  et~al., 2021, \mn@doi [Astrophys. J.] {10.3847/1538-4357/abe40e},
  910, 152

\bibitem[\protect\citeauthoryear{de Lera~Acedo et~al.}{de~Lera~Acedo
  et~al.}{2022}]{deLeraAcedo2022}
de Lera~Acedo E.,  et~al., 2022, \mn@doi [Nature Astron.]
  {10.1038/s41550-022-01817-6}, 6, 998

\bibitem[\protect\citeauthoryear{van~der Walt, Colbert  \& {Varoquaux}}{van~der
  Walt et~al.}{2011}]{vanderWalt:2011bqk}
van~der Walt S.,  Colbert S.~C.,   {Varoquaux} G.,  2011, \mn@doi [Comput. Sci.
  Eng.] {10.1109/MCSE.2011.37}, 13, 22

\makeatother
\end{thebibliography}



\appendix

\section{Hierarchical Inference}\label{app: hier}
Let $\boldsymbol{\Lambda}$ be a set of hyperparameters that determines the shape of the model distribution of some parameter $\theta$ in an observation. The likelihood for the data $d$ given hyperparameters $\Lambda$ can be written as~\citep{Thrane:2018qnx, Talbot:2017yur}
\begin{equation}
    \label{eq: lik_d_lamda_app}
    \mathcal{L}(d \vert \boldsymbol{\Lambda}) = \int d\theta \mathcal{L}(d \vert \theta) \pi(\theta \vert \boldsymbol{\boldsymbol{\Lambda}})\,,
\end{equation}
where $\mathcal{L}(d \vert \theta)$ is the likelihood for getting the data $d$ given $\theta$ and $\pi(\theta \vert \boldsymbol{\Lambda})$ is the conditional prior on $\theta$. Using Bayes theorem, we can write the hyper-posterior as
\begin{equation}
    \label{eq: hyper_post_app}
    p(\boldsymbol{\Lambda} \vert d) = \frac{\mathcal{L}(d \vert \boldsymbol{\Lambda}) \pi(\Lambda)}{\mathcal{Z}_{\Lambda}}
\end{equation}
where $\mathcal{Z}_{\Lambda} \equiv \int d\boldsymbol{\Lambda} \mathcal{L}(d \vert \boldsymbol{\Lambda}) \pi(\boldsymbol{\Lambda})$ is the hyper-evidence. 

For $N$ independent events, we can write $\mathcal{L}(\boldsymbol{d} \vert \theta) = \prod_i^N \mathcal{L}(d_i \vert \theta)$, where $d_i$ is the data of the $i$th event. 
Assuming that all $N$ events are coming from the same population model distribution, we can write the total hyper-likelihood as:
\begin{equation}
    \label{eq: tot_hlike_app}
    \mathcal{L}_{\rm tot}(\boldsymbol{d} \vert \boldsymbol{\Lambda}) = \prod_i^N \mathcal{L}(d_i \vert \boldsymbol{\Lambda}) =  \prod_i^N \int d\theta \mathcal{L}(d_i \vert \theta) \pi(\theta \vert \boldsymbol{\Lambda}),
\end{equation}
while the hyper-posterior takes the form:
\begin{equation}
    \label{eq: tot_hyper_post_app}
    p_{\rm tot}(\boldsymbol{\Lambda} \vert \boldsymbol{d}) = \frac{\mathcal{L}_{\rm tot}(\boldsymbol{d} \vert \boldsymbol{\Lambda}) \pi(\boldsymbol{\Lambda})}{\mathcal{Z}_{\boldsymbol{\Lambda}, \rm tot}}.
\end{equation}
\section{Fisher Matrix Analysis}\label{app: fisher}
Let $h(\boldsymbol{\Theta};t)$ be the GW signal emanated from a CBC. Then the Fisher matrix $\bf \Gamma$ is defined as
\citep{1994PhRvD..49.2658C}:
\begin{equation}
	\Gamma_{ij} = \left (\frac{\partial h}{\partial \Theta_i} \Bigg \vert \frac{\partial h}{\partial \Theta_j} \right)
\end{equation}
where $\Theta_{i,j}$ are the CBC's intrinsic and extrinsic parameters that determine the shape of the GWs. $(|)$ represents a noise-weighted inner product of two time series $a(t)$ and $b(t)$ and is defined as:
\begin{equation}
	(a | b) = 4 \mathfrak{R} \int_{f_\mathrm{min}}^{f_\mathrm{max}} \dfrac{\tilde{a}^*(f) \tilde{b}(f)}{S_n(f)} df
\end{equation}
where $\tilde{a}(f)$ and $\tilde{b}(f)$ are the Fourier Transforms of $a$ and $b$, respectively, $*$ represents the complex conjugate of the Fourier Transforms, $f$ is the GW frequency, $f_{\rm min}$ and $f_{\rm max}$ are the minimum and maximum frequencies of the signal, respectively, and $S_n(f)$ is the Power Spectral Density (PSD) of the GW detector. The inverse of Fisher Matrix  $\boldsymbol{\Sigma}$ is called the Covariance Matrix and contains the information about the errors in the measurement of $\boldsymbol{\Theta}$. Specifically, the square roots of the diagonal components of $\boldsymbol{\Sigma}$, i.e., $\sqrt{\Sigma^{ii}}$, provide the r.m.s. errors on $\Theta_i$.

\section{Additional Plots}\label{app: additional_plots}
\begin{figure}
    \centering
    \includegraphics[width=0.985\linewidth]{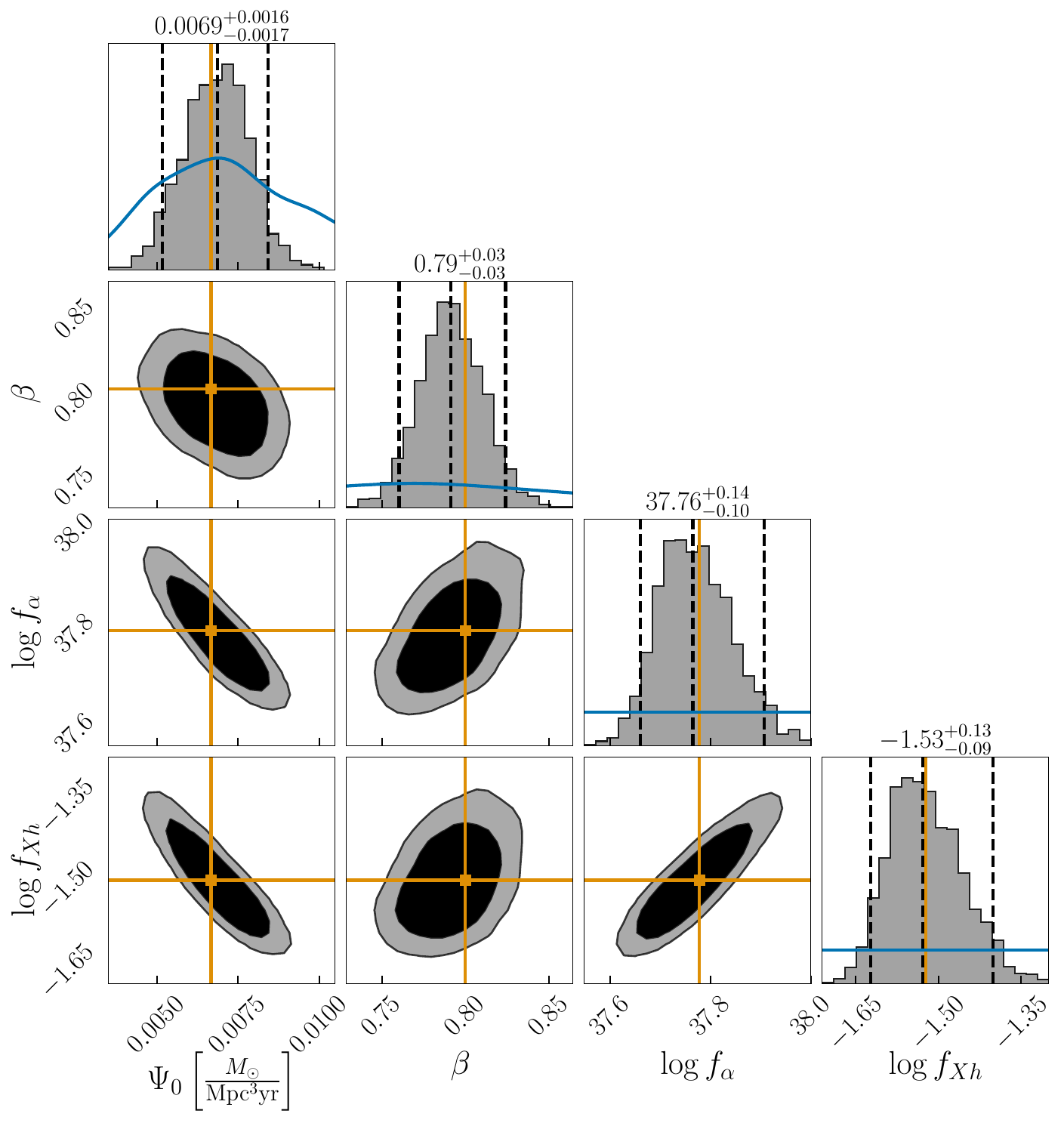}
    \vspace{-10pt}
    \caption{The inferred posteriors of 21-cm global signal parameters obtained using the GW hierarchical inference posteriors as priors (see Figure~\ref{fig: gw_post}). The 100 GW-events case is assumed here. The black dashed line represents the $5th$, $50th$, and $95th$ percentiles, the orange lines represent the true values, the blue lines represent the priors, while the $2d$-contours correspond to $68\%$ and $90\%$ credible regions.}
    \label{fig: 21cm_corner_plot}
\end{figure}

\begin{figure}
    \centering
    \includegraphics[width=0.985\linewidth]{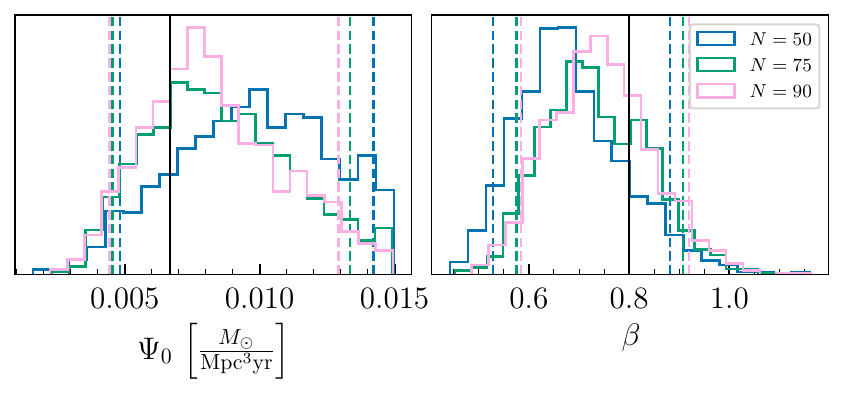}
    \vspace{-10pt}
    \caption{The inferred posteriors of SFRD parameters obtained using the GW hierarchical inference. The colors of the posteriors correspond to the number of BBH events used in inference, and the corresponding dashed lines represent the 90\% credible regions.}
    \label{fig: varying_nev}
\end{figure}

Figure~\ref{fig: 21cm_corner_plot} shows a corner plot of the posteriors pertaining to the inference of the 21-cm global signal parameters. The priors on the $\Psi_0$ and $\beta$ parameters are taken to be the posteriors acquired from the hierarchical inference using $100$ BBH (GW) events. In addition, we also show histograms of SFRD parameters for different numbers of BBHs in Figure~\ref{fig: varying_nev} to show how many events would suffice to constrain both SFRD parameters simultaneously.

\label{lastpage}
\end{document}